\documentclass[%
 jmp,
 pre,%
 amsmath,amssymb,
 reprint,%
author-year,%
author-numerical,%
showpacs
]{revtex4-1}
\usepackage[dvipdfmx]{graphicx}
\usepackage{dcolumn}
\usepackage{bm}
\usepackage[dvips]{color}

\def\vc#1{\mbox{\boldmath $#1$}}

\def\abs#1{\left|#1\right|}

\def\lb{\left(}
\def\rb{\right)}

\def\beq{\begin{eqnarray}}
\def\eeq{\end{eqnarray}}

\def\dps{\displaystyle}

\begin{document}


\title[Dynamical scaling of fragment distribution in drying paste]{Dynamical scaling of fragment distribution in drying paste}

\author{Shin-ichi Ito}
\author{Satoshi Yukawa}
\affiliation{Department of Earth and Space Science, Graduate School of Science, Osaka University, Toyonaka 560-0043, Osaka, Japan}

\date{\today}
\begin{abstract}
We reproduce patterns of drying paste by means of smoothed particle hydrodynamics which is the one of methods for solving the equations of continuum in the Lagrangian description.
In addition to reproduce a realistic pattern, we find that average size of fragments decays in proportion to inverse time in the case of a linear drying process. Distributions of the size of the fragments are obtained depending on the time. 
We find a universal scaling distribution by scaling analysis with the average size of the fragment. 
\end{abstract}


\maketitle

\section{\label{sec:level1}Introduction}
We often see patterns of fracture on drying lakes, paddy fields and so on.
It is well-known that fractures of drying paste show different properties depending on the thickness of   the paste.
The drying cracks are classified into two types.
In the case that the thickness is larger than 
a horizontal length of system, 
the fragment's characteristic size on upper surface is limited
and the cracks run slowly along depth direction. In this case, the cracks make prismatic structures. 
On the other hand, in the case that thickness is shorter than the size of system, the cracks approach the bottom of container immediately and do not make columnar structures.
In this case, some interesting properties have been reported.
Groisman\cite{A.Groisman1994} did some experiments by using coffee granular and reported that the average size of fragments is proportional to the thickness of the paste. 
Furthermore, Nakahara and Matsuo\cite{Nakahara2005,Nakahara2006} reported that some kind of paste remembers the force they received or the directions they flowed before drying on their crack patterns.
Thus the properties of drying paste have been investigated vigorously,
however, most of them were investigated only at the time when drying has finished. 
The patterns change every moment until the end of drying and the statistical properties change every moment.
It is worth investigating the dynamical and statistical properties for understanding fractures of drying paste.
There are studies for velocity of crack tips\cite{Kitsunezaki2009} as the time-dependent properties, 
however, there are few studies of how patterns change with time.
In this article, we focus on the properties of size of fragment as the time-dependent properties.

Smoothed Particle Hydrodynamics(SPH)\cite{Gingold1977,L.B.Lucy2004} is the one of methods to calculate equations of continuum. SPH has been developed in the field of astrophysics to solve problems of compressible flow\cite{Monaghan1988,Benz1988}. 
Currently, SPH has been applied to calculations of incompressible fluids\cite{Morris1997,Hu2007} and, furthermore,   elastic\cite{Gray2001} or visco-elastic materials\cite{Fang2006}.
There are also some studies of plastic-elastic materials\cite{Bui2008,Libersky1993} and the formalization of treatments for ductile materials is nearly complete, however, the formalization of treatments for brittle materials have not been enough yet.

SPH belongs to a method in the Lagrangian description.
In the Lagrangian description, we do not have to make meshes. 
Therefore it is easy to calculate the equations of systems which have complex free surfaces.
In the case of drying fracture process, new free surfaces are created when cracks run.
As a consequence, complex free surfaces are created.
Therefore, SPH is a suitable method for simulations of drying paste.
However, there is a problem in the implementation of brittle fractures in the SPH algorithm. 
Resolving this problem is the key for reproducing fractures of drying paste by using SPH.

The aim of this article are (i) to reproduce a crack pattern of a thin drying paste with a continuum model by giving the implementation of the brittle material in the SPH algorithm, 
and (ii) to investigate the dynamical and statistical properties of size of fragment with the SPH simulation. 
To simplify the calculation, we make a two-dimensional continuum model from a three-dimensional Voigt visco-elastic continuum model.
We investigate two characteristic quantities which are convenient for understanding time evolution of the fragments size:
the average size and the size distribution.
As a result, we find that the average size decays in proportion to inverse time in the case of a linear drying process. 
In addition, we find a universal time-independent scaling distribution by the scaling analysis with the average size of the fragments.  

\section{\label{sec:level3}Model of drying paste}
\begin{figure}[tbp]	
	\begin{center}
	\includegraphics[width=0.9\linewidth]{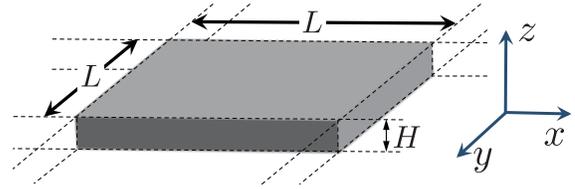}
	\end{center}
	\caption{Geometry of the paste we simulate in this paper. Thickness is $H$ and length of a side is $L$.}
	\label{fig:paste}
\end{figure}
In this paper, we take a thin layer of paste whose thickness is ${H}$ and length of a side is ${L}$ (See Fig.~\ref{fig:paste}).
Let ${\vc v}$, ${\vc u}$, ${\vc \sigma}$ and ${\rho}$ denote velocity, displacement, stress and density, respectively. 
In general, the equations of motion, continuity and displacement of three dimensional continuum are described as follows:
\begin{eqnarray}
	\rho \frac{d \vc v}{dt}=\nabla \cdot \vc \sigma,\label{eq:eq1}
\end{eqnarray}
\begin{eqnarray}
	\frac{d\rho}{dt}=-\rho\nabla \cdot \vc v,\label{eq:eq2}
\end{eqnarray}
\begin{eqnarray}
	\frac{d\vc u}{dt}=\vc v.\label{eq:eq3}
\end{eqnarray}
Here, ${d/dt}$ is a time differential operator in the Lagrangian description.
In this paper, we treat the paste as the Voigt visco-elastic material.
In addition, we take the effect of drying into account.
The equation of stress ${\vc \sigma}$ is given by
\begin{eqnarray}
	\vc \sigma = \vc \sigma_{vis} +\vc \sigma_{el} +\vc \sigma_{dry}.\label{eq:eq4}
\end{eqnarray}
Here, ${\vc \sigma_{vis}}$ is a viscous part. By using a strain velocity ${\dot{\vc \epsilon}}$ given by
\begin{eqnarray}
	\dot{\vc \epsilon}=\frac{1}{2}\left\{(\nabla \otimes \vc v)+(\nabla \otimes \vc v)^{T}\right\},\label{eq:eq5}
\end{eqnarray}
${\vc \sigma_{vis}}$ is described as
\begin{eqnarray}
	\vc \sigma_{vis}=\eta\dot{\vc \epsilon},\label{eq:eq6}
\end{eqnarray}
where ${\eta}$ is a viscosity coefficient.
${\vc \sigma_{el}}$ is an elastic part. By using a rotation velocity ${\vc \Omega}$ given by
\begin{eqnarray}
	\vc \Omega=\frac{1}{2}\left\{(\nabla \otimes \vc v)-(\nabla \otimes \vc v)^{T}\right\},\label{eq:eq7}
\end{eqnarray}
the time evolution of ${\vc \sigma_{el}}$ is described as
\begin{eqnarray}
		\frac{d\vc \sigma_{el}}{dt}=
		                  \lambda
 \mbox{\rm Tr}\left(\dot{\vc \epsilon}\right)\vc I 
	              +2\mu
\dot{\vc \epsilon} 
	              + \vc \sigma_{el} \cdot \vc \Omega 
	              -\vc \Omega \cdot \vc \sigma_{el},\label{eq:eq8}
\end{eqnarray}
where ${\lambda}$ and ${\mu}$ are Lame's elastic constants.
${\vc I}$ is a unit tensor.

${\vc \sigma_{dry}}$ is a drying part. 
Drying makes inner stress increased with time evolution.
In general, we do not know the functional form of ${\vc \sigma_{dry}}$:
The stress is considered as the monotone increasing function of time ${t}$ in initial stage of drying process.
Therefore, we take ${\vc \sigma_{dry}}$ as a linear function of ${t}$.
Furthermore, we assume that the effect of drying only appears in diagonal part of ${\vc \sigma_{dry}}$.
In a previous work\cite{Kitsunezaki2004}, the drying part was treated similarly to the present form. 
Eventually, ${\vc \sigma_{dry}}$ is described by
\begin{eqnarray}
	\vc \sigma_{dry}=v_{\tau}t\vc I,\label{eq:eq9}
\end{eqnarray}
where ${v_{\tau}}$ is the drying speed.

To simplify the calculation, we approximate the equation of motion to 
the two-dimensionalized one.
In order to approximate, we use the discretization used in the study by Otsuki\cite{Otsuki2005}. 
Let lower index ${i}$ denote that the quantity is along the ${i}$ direction.
${u_{i}}$, ${v_{i}}$ at ${\lb x,y,0\rb}$ and ${\sigma_{iz}}$ at ${\lb x,y,H\rb}$ must be satisfied boundary conditions as follows:
\begin{eqnarray}
	\sigma_{iz}\left(x,y,H\right)=0,\label{eq:ap1} &\\
	u_{i}\left(x,y,0\right)=0. \label{eq:ap2}
\end{eqnarray}
If the thickness of paste ${H}$ is thin enough, we can ignore the motions along ${z}$ direction and consider that the space differential of a certain quantity along ${z}$ direction is approximated to the difference between the quantities at the top (${z=H}$) and the bottom (${z=0}$) of the paste.
For example, let ${\epsilon_{iz}}$ denote the strain along ${z}$ direction and it at ${\lb x,y,0\rb}$ is given by  
\begin{eqnarray}
	\epsilon_{iz}\left(x,y,0\right)&=&
	\frac{1}{2}\left(\frac{\partial u_{i}}{\partial z}\left(x,y,0\right)+\frac{\partial u_{z}}{\partial x_{i}}\left(x,y,0\right)\right) \nonumber\\
	&\simeq&\frac{1}{2}\frac{u_{i}\left(x,y,H\right)-u_{i}\left(x,y,0\right)}{H} \nonumber\\
	&=&\frac{1}{2}\frac{u_{i}\left(x,y,H\right)}{H}
	\enspace.\label{eq:ap3}
\end{eqnarray}
${\dot{\epsilon}_{iz}\left(x,y,0\right)}$ is also descritized similarly. 
When we consider the motion of the upper surface of the paste, we can divide the right-hand side of Eq.~\eqref{eq:eq1} into the following equations:
\beq
	\rho \frac{d v_{i}}{dt}=\partial_{j}\sigma_{ij} +\frac{\partial \sigma_{iz}}{\partial z} \quad 
	\text{for $i,j= x, y$ at ${\lb x,y,H\rb}.$}\label{eq:ap5}
\eeq
The second term in the right-hand side of Eq.~\eqref{eq:ap5} works as a resistance force for the motion along ${i}$ direction.
This resistance force can be discretized as the following form:
\begin{eqnarray}
	\frac{\partial \sigma_{iz}}{\partial z} \left(x,y,H\right)&\simeq&
	\frac{\sigma_{iz}\left(x,y,H\right)-\sigma_{iz}\left(x,y,0\right)}{H}\nonumber\\	
	&=&-\frac{\sigma_{iz}\left(x,y,0\right)}{H} 
	\nonumber\\
	&=&-2\mu \frac{\epsilon_{iz}\left(x,y,0\right)}{H} -\eta \frac{\dot{\epsilon_{iz}}\left(x,y,0\right)}{H} \nonumber\\
	&=&	-\frac{\mu}{H^{2}}u_{i}\left(x,y,H\right)-\frac{\eta}{2H^{2}}v_{i}\left(x,y,H\right)
	\enspace.\nonumber\\ \label{eq:eqvv}
\end{eqnarray}
Here, we suppose the damping force, the second term of the right-hand side of Eq.~\eqref{eq:eqvv}, 
can be ignored because this term is effective only just in the time when cracks break out and is very smaller than the first term of the right-hand side of Eq.~\eqref{eq:eqvv} if the system shrinks slowly.
Eventually, the two dimensional equation of motion is given by 
\begin{eqnarray}
	\rho \frac{d \vc v}{dt}=\nabla \cdot \vc \sigma - \frac{\mu}{H^{2}}\vc u
	\enspace , \label{eq:eq10}	
\end{eqnarray}
where ${\vc v}$, ${\vc u}$ and ${\vc \sigma}$ are two dimensional quantities, respectively.
We calculate the motion of the upper surface by using Eqs.~\eqref{eq:eq2}${\sim}$\eqref{eq:eq9} and \eqref{eq:eq10}.

When we simulate fractures of continuum, we have to set a yield criterion.
There are many yield criterions which depend on the kind of simulated material.
In this paper, we use a criterion\cite{hoover2006smooth}  by using local averaged stress ${\bar{\sigma}=\mbox{\rm Tr}\left(\vc \sigma \right)/2}$, which corresponds to the pressure.
Local averaged stress criterion is described as follows:
(1) Calculate ${\bar{\sigma}}$ at all positions in the material.
(2) If ${\bar{\sigma}}$ is grater than a definite yield stress ${\sigma_{Y}}$, we make the stress at the position into zero.	

\section{\label{sec:level4}smoothed particle hydrodynamics and simulation condition}
SPH\cite{L.B.Lucy2004,Gingold1977} is the method for solving the equations of continuum by using a movable mesh point ``particle''. All particles 
have the physical quantities.
Let upper indexes written with capital letters denote particle numbers.  
Let ${\vc r^{I}}$ denote the position of particle ${I}$.
 In SPH formula, a general physical quantity ${f^{I}}$ is given by
\begin{eqnarray}
	f^{I}=\sum_{J}f^{J}W\left(\left|\vc r^{I}-\vc r^{J}\right|;h\right)\frac{m^{J}}{\rho^{J}},\label{eq:sphdis}
\end{eqnarray}
where ${m^{J}}$ and ${\rho^{J}}$ are the mass and density of particle ${J}$.
The summation of ${J}$ is calculated by using all particles in the system.
The function ${W(x;h)}$ is a kernel function and has a positive parameter ${h}$ called an ``effective length''.
${W(x;h)}$ is required to be the Dirac delta function under the limit of ${h\searrow0}$.
The accuracy of the simulation is related with the choice of the kernel function $W(x;h)$.
In previous works\cite{Gray2001} of SPH simulation, a spline function is often used.
In this paper, we choose the fifth order spline function as the kernel function as follows:
\begin{equation}
	W(x;h)=\frac{63}{478\pi h^{2}}W_{5}\left(\frac{x}{h}\right) \enspace ,
\end{equation}
where ${W_{5}\left(x\right)}$ is the fifth order spline function:
\begin{align}
W_{5}(x)&=
\begin{cases}
	q_{3}\left(x\right)-6q_{2}\left(x\right)+15q_{1}\left(x\right) & (\abs x \le \frac{1}{3}) \\[1mm]
	q_{3}\left(x\right)-6q_{2}\left(x\right)                                & (\frac{1}{3} < \abs x \le \frac{2}{3}) \\[1mm]
	q_{3}\left(x\right)                                                           & (\frac{2}{3} < \abs x \le 1)\\[1mm]
	0                                                                & (1 < \abs x),
\end{cases}\nonumber 
\end{align}
with 
\begin{align}
q_{k}\left(x\right)&=\left(k-3\abs x \right)^{5}\quad \left(k=1,2,3\right).\nonumber
\end{align}

In SPH description, a gradient of the physical quantity ${\nabla f^{I}}$ is given by
\begin{eqnarray}
	\nabla f^{I}=\sum_{J}f^{J}\nabla W\left(\left|\vc r^{I}-\vc r^{J}\right|;h\right)\frac{m^{J}}{\rho^{J}}.\label{eq:sp1}
\end{eqnarray}

By using Eq.~\eqref{eq:sp1} and some kind of differential transportation, ${\nabla \cdot \vc \sigma}$, ${\nabla \otimes \vc v}$ and ${\nabla \cdot \vc v}$ are discretized as follows:
\begin{align}
\left(\frac{\nabla \cdot \vc \sigma}{\rho}\right)^{I}&=
\sum_{J}m^{J}\nabla W^{IJ}\cdot \left\{\frac{\vc \sigma^{I}}{(\rho^{I})^{2}}+\frac{ \vc \sigma^{J}}{(\rho^{J})^{2}}\right\} ,\label{eq:sp2} \\
\left(\nabla \otimes \vc v\right)^{I}&=
\sum_{J}\frac{m^{J}}{\rho^{I}}\nabla W^{IJ}\otimes\left(\vc v^{J}-\vc v^{I}\right), \label{eq:sp3} \\
\left(\nabla \cdot \vc v\right)^{I} &=\mathrm{Tr}\left(\left(\nabla \otimes \vc v\right)^{I}\right), \label{eq:sp33}
\end{align}
where ${\nabla W^{IJ}}$ denotes ${\nabla W\left(\left|\vc r^{I}-\vc r^{J}\right|;h\right)}$.

When we calculate the time evolutions of Eqs.~\eqref{eq:eq2}, \eqref{eq:eq3}, \eqref{eq:eq8} and \eqref{eq:eq10} in SPH description, we do not have to calculate advection term since SPH is based on the Lagrangian description. 
Instead of calculating the advection term, we have to calculate the time evolutions of positions of particles as the following:
\begin{eqnarray}
	\frac{d\vc r^{I}}{dt}=\vc v^{I}.\label{eq:sp4} %
\end{eqnarray}

To avoid numerical instability, we use a velocity averaging method.
In SPH simulations, 
interaction between particles has no repulsive core. 
Therefore, in extreme conditions, there are some possibilities of that particles close each other. It causes numerical instability of the calculation.  
To avoid this instability, the velocity averaging method is proposed by Monaghan\cite{Monaghan1989,Gray2001}.
According to the Monaghan's description, the velocity of particle ${I}$ is averaged as the following:
\begin{eqnarray}
	\vc v^{I} \rightarrow \vc v^{I} +\tilde{\epsilon} \sum_{J}\frac{m^{J}}{\rho^{IJ}}\left(\vc v^{J}-\vc v^{I}\right)W\left(\left|\vc r^{I}-\vc r^{J}\right|;h\right),\nonumber\\ \label{eq:eq11}
\end{eqnarray}
where ${\rho^{IJ}}$ denotes ${\frac{1}{2}\left(\rho^{I}+\rho^{J}\right)}$ and ${\tilde{\epsilon}}$ is a tuning parameter.
In this paper, we choose ${\tilde{\epsilon}=0.5}$. 

In the following study, Eqs.~\eqref{eq:eq2}${\sim}$\eqref{eq:eq9} and \eqref{eq:eq10} discretized by using Eqs.~\eqref{eq:sp2}${\sim}$\eqref{eq:sp4} and \eqref{eq:eq11} are taken to be the basic equations of the model. 

We have to apply the local average stress criterion to SPH.
It is plausible that fractures of drying paste are treated as brittle fractures.
In simple consideration, we only have to make the stress of particle into zero, when the stress becomes grater than the yield stress ${\sigma_{Y}}$.
However, we can not reproduce brittle fractures in this way because stress can be relaxed rapidly; 
The stress on the surfaces of brittle cracks is relaxed as long as both the surfaces are within the length ${h}$.
In SPH description, there are some different treatments of brittle fractures\cite{hoover2006smooth,Benz1995}.
In this paper, we achieve the brittle fracture by removing particles which satisfy local average stress criterion 
instead of resetting the stress. 
Using this method, the distance between two created surfaces 
becomes enough large which prevents the stress relaxation.
\begin{table}[tb]	
	\begin{tabular}{lcc}
	\hline
	\hline
	Parameter & Symbol & Value\\ \hline
	First Lame's constant&${\lambda}$          & 1.0\\
	Second Lame's constant&${\mu}$     & 0.1 \\
	Yield stress&${\sigma_{Y}}$ & 5.0${\times 10 ^{-3}}$\\
	Viscosity&${\eta}$ & 1.0\\
	Thickness of paste&${H}$ &0.316\\
	Drying speed&${v_{\tau}}$ &2.2${\times 10^{-5}}$\\
	Length of a side&${L}$ &10.0\\
	Effective length&${h}$ &0.2\\
	Number of SPH particles&${\rm N}$ &4.0${\times 10^{4}}$ \\
	Intial density&${\rho_{0}}$ & 1.0\\
	Intial stress&${\vc \sigma_{0}}$ & -0.01${\sim}$0.01${\times \sigma_{Y}}$\\
	\hline
	\hline
	\end{tabular}
	\caption{\label{tab:para}Parameters and initial quantities used in the simulation.}
\end{table}

In the actual simulation, we prepare a square paste whose length of a side is ${L}$.
We impose a periodic boundary condition on the paste.
The values of parameters and initial physical quantities are summarized in TABLE \ref{tab:para}.
These values are non-dimensionalized by ${\rho_{0}}$, ${\lambda}$ and ${\eta}$.
The unit of time and space are given by ${\eta/\lambda}$ and ${\eta/\sqrt{\rho_{0}\lambda}}$.
Initial positions of particles are located randomly to avoid being created anisotropic patterns.
Initial displacement and velocity are taken to be zero.
Initial density ${\rho_{0}}$ is set the value shown in TABLE \ref{tab:para}, which is also the unit.
Initial stress ${\vc \sigma_{0}}$ is randomly chosen from the uniform distribution shown in TABLE~\ref{tab:para}. 
The mass ${m}$ of the particle is calculated by using the initial density ${\rho_{0}}$ and the initial position ${\vc r}$ by using the consistent condition:
\beq
	\rho_{0}=\sum_{J}m^{J}W\lb \left| \vc r^{I} - \vc r^{J} \right|;h\rb.
	\label{eq:calmass}
\eeq
Equation~\eqref{eq:calmass} is linear simultaneous equations and can be solved as a large sparse matrix problem.
In order to solve Eq.~\eqref{eq:calmass}, we use the conjugate gradient method.
We calculate the time evolutions of quantities by using the fourth order Runge-Kutta (RK4) and Predictor-Corrector (PC4) methods. 
RK4 is used both in a few initial time steps and in several time steps after removals of particles
and PC4 is used in otherwise time steps. 
This is because the physical quantities become discontinuous when removals of particles occur 
and the calculation may become unstable if we only use PC4.

\section{\label{sec:level5}Results}

\begin{figure}[t]	
	\begin{center}
	\includegraphics[width=1.0\linewidth]{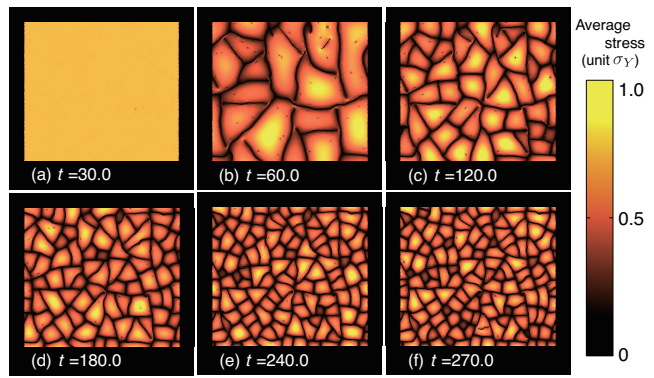}
	\caption{\label{fig:snapshots}Snapshots of time evolution of drying model simulated with parameters shown in TABLE \ref{tab:para}. 
	 From the upper-left figure, time passes in alphabetical order.
	 Color shows magnitude of average stress. Black region is the crack. 
	  }	
	\end{center}
\end{figure}

\subsection{Time evolution of a crack pattern}

Figure~\ref{fig:snapshots} shows the snapshots of a simulation of drying model.
From the upper-left figure, time passes in alphabetical order.
In this figure, we can see some nucleations of cracks before the cracks run.
There are some cases where the cracks do not grow.
We have not understood the reason why there are cases whether cracks run or not.
Cracks run almost straight. 
When a crack runs against into another one, they cross each other normally. 
As a result, fragments become almost convex polygonal shape.
In the simulation, crack tips' velocities are the fastest at the time when the cracks break out and decay with approaching to another one. 
These behaviors are very similar to real patterns of drying paste.

Color shows a magnitude of the average stress normalized by the yield stress.
In this figure, we can see that the average stress is almost zero along the edges of each fragments 
and becomes to maximum on the position near the center of area.
We can consider that this is because fragments are almost convex polygonal shape.
Since a fragment shrinks isotropically, if the fragment is convex polygonal shape,
stress concentration tends to occur near the center of area and rarely occurs along the edges.
 
\subsection{\label{sec:lvel6}Time evolution of average size of fragments}

\begin{figure}[tbp]	
	\begin{center}
	\includegraphics[width=1.0\linewidth]{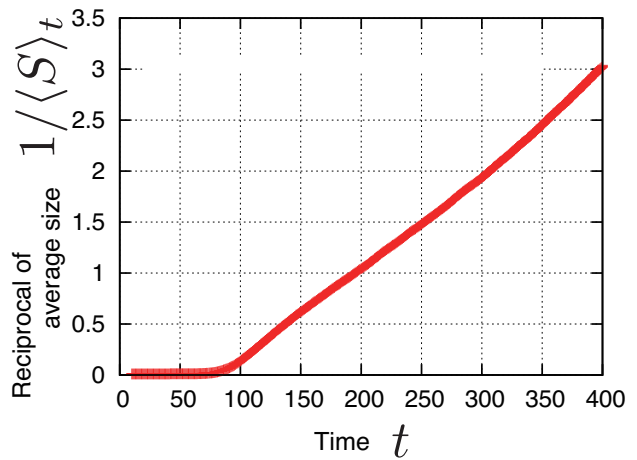}
	\caption{\label{fig:average}Time evolution of average size ${\langle S\rangle_{t}}$: Vertical axis indicates reciprocal of ${\langle S\rangle_{t}}$. Horizontal axis indicates the time.  
	}
	\end{center}	
\end{figure}

In order to investigate the time evolution of average size of fragments, 
we prepare a hundred samples with a different initial stress.
To calculate each area of fragments, we binarize the snapshots as following way:
We discretized the snapshot specially with a square mesh. 
On each mesh point ${\vc x}$, a binary value ${\phi\left(\vc x\right)}$ is calculated by
\beq
\phi\lb\vc x\rb=
	\begin{cases}
		1 & \text{for } \phi_{0} \le \sum_{J}\frac{\dps m^{J}}{\dps \rho^{J}}W\lb\vc x -\vc r^{J};h\rb\\
		0 & \text{otherwise,}
\end{cases}
\nonumber\eeq
where $\phi_{0}$ is a threshold value.
If enough particles surround the mesh point ${\vc x}$,  ${\phi\lb\vc x\rb}$ becomes ${1}$.
This means that the position is the inside of a fragment.
Conversely, if few particles surround the mesh point ${\vc x}$, ${\phi\lb\vc x\rb}$ becomes ${0}$.
This means that the position is the outside of a fragment and a part of cracks.
We can calculate the area of fragment by clustering and summing up the value of ${\phi}$. 
In this paper, we choose ${\phi_{0}=0.8}$ and the interval of square mesh ${\Delta x=0.01h}$.

\begin{figure}[t]	
	\begin{center}
	\includegraphics[width=1.0\linewidth]{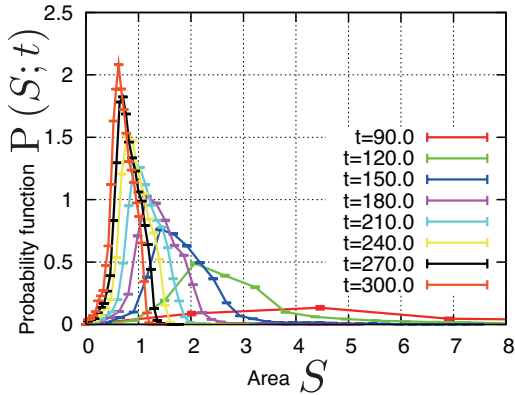}
	\caption{\label{fig:hist1}Size distributions of fragments: Horizontal axis indicates size of fragment and vertical axis  indicates the distribution. The distributions from ${t=90.0}$ to ${t=300.0}$ are shown in this figure.}
	\end{center}	
\end{figure}

Let $\langle S\rangle_{t}$ denote the average size of fragments at time ${t}$. Figure~\ref{fig:average} shows the time evolution of $\langle S \rangle_{t}$.
As we see in this figure, the reciprocal value of $\langle S\rangle_{t}$ evolves in proportion to time except for the initial stage.
In other words, ${\langle S\rangle_{t}}$ evolves in inverse proportion to time in long time.
In the initial stage, the time which is earlier than ${t\sim150}$, 
we find that 
there are some cases of that an initial fragment have not broken into two or more pieces yet 
and ones of that an initial one have already broken.

This property can be explained by means of an dimensional analysis. 
We consider an over-damped equation of motion which is made by ignoring the inertial term of Eq.~\eqref{eq:eq10}. The equation is given by 
\beq
	\frac{\partial \sigma_{ij}}{\partial x_{j}}=\frac{\mu}{H^{2}} u_{i}.\label{eq:odeq1}
\eeq
Furthermore, the constitutive equation ignoring the viscous term of Eq.~\eqref{eq:eq4} and using a general drying stress ${F\lb t \rb}$ is given by
\beq
	\sigma_{ij}=\left\{ \lambda \frac{\partial u_{k}}{\partial x_{k}} + F\lb t \rb \right\} I_{ij} +\mu \lb \frac{\partial u_{i}}{\partial x_{j}} +\frac{\partial u_{j}}{\partial x_{i}} \rb.\label{eq:odeq2}
\eeq
We investigate Eqs.~\eqref{eq:odeq1} and \eqref{eq:odeq2} by means of dimensional analysis.
Let ${U\lb t\rb}$, ${\Xi\lb t\rb}$ and ${L\lb t\rb}$ denote a characteristic displacement, stress and length of the fragment, respectively.
Note that the space differential operator ${\partial/\partial x}$ is replaced into ${1/L\lb t\rb}$. 
By using these characteristic quantities, Eqs.~\eqref{eq:odeq1} and \eqref{eq:odeq2} are rewritten as follows:
\beq
	-\frac{\Xi\lb t\rb}{L\lb t\rb}=\frac{\mu}{H^{2}} U\lb t\rb,
\eeq
\beq
	\Xi\lb t\rb=\lb \lambda+2\mu\rb \frac{U\lb t\rb}{L\lb t\rb} + F\lb t\rb.
\eeq
From these equations, we obtain the relationship between ${\Xi\lb t\rb}$, ${L\lb t\rb}$ and ${F\lb t\rb}$ 
as 
\beq
	L^{2}\lb t\rb=\frac{\lambda+2\mu}{\mu}\frac{H^{2}}{F\lb t\rb/\Xi\lb t\rb -1}.\label{eq:cleq} 
\eeq
The value of ${\Xi\lb t\rb}$ is limited by the yield stress.
Therefore the time evolution of ${\Xi\lb t\rb}$ is ignored from the time evolution of ${L^{2}\lb t\rb}$.
Average size ${\langle S\rangle_{t}}$  must be corresponding to ${L^{2}\lb t\rb}$ which means a characteristic area.
Eventually, ${\langle S\rangle_{t}}$ is proportional to ${1/F\lb t\rb}$.
Therefore ${\langle S\rangle_{t}}$ is proportional to ${1/t}$ because ${F\lb t\rb}$ is proportional to ${t}$ in our case.  
As a practical matter, ${\langle S\rangle_{t}}$ must be described by superposition of various ${L^{2}\lb t\rb}$, however, it does not affect the time dependence of ${\langle S\rangle_{t}}$ in long time. 

\subsection{\label{sec:lvel7}Time evolution of size distributions of fragments}
\begin{figure}[t]	
	\begin{center}
	\includegraphics[width=1.0\linewidth]{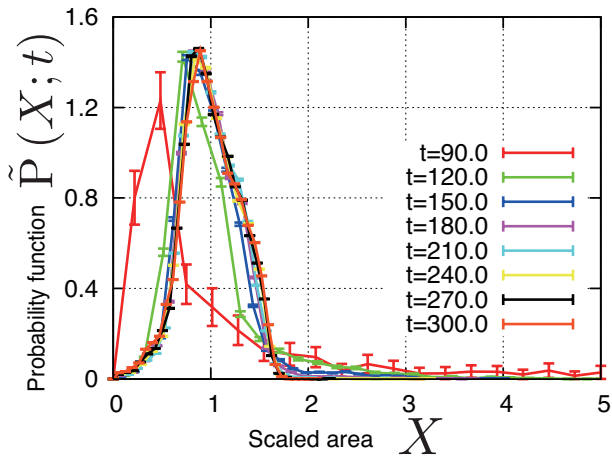}
	\caption{\label{fig:hist2}Size distributions scaled by their averages: Horizontal axis indicates scaled size $X$. Vertical axis indicates the scaled distribution. The distributions from ${t=90.0}$ to ${t=300.0}$ are shown in this figure.}
	\end{center}	
\end{figure}

Next we investigate the time evolution of size distributions of fragments.
Let $\mathrm{P}\left(S;t\right)$ denote the size distribution at time ${t}$. Figure~\ref{fig:hist1} shows the results.
As we can see in this figure, the peak of distributions goes to smaller side with time evolution. This shift is trivial behavior.  
In the case of the present model, this peak will go to zero in the long-time limit since the end of drying is not taken into the account in this model. 

\begin{figure}[t]	
	\begin{center}
	\includegraphics[width=1.0\linewidth]{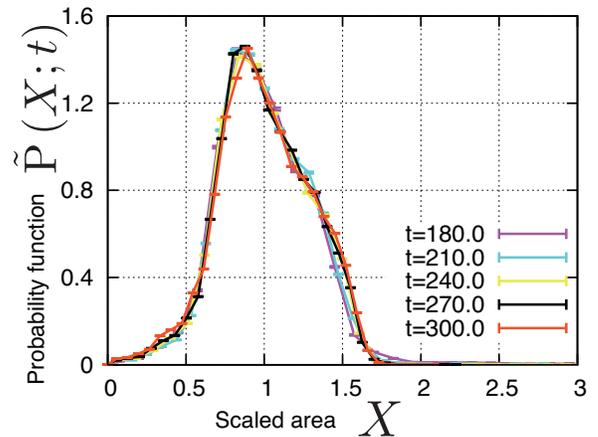}
	\caption{\label{fig:hist3}This figure shows the scaled distributions of FIG.\ref{fig:hist2} from ${t=180.0}$ to ${t=300.0}$. }
	\end{center}	
\end{figure}

As a property of distributions, we find that the distributions can be scaled by their averages (see Fig.~\ref{fig:hist2} and the details in Fig.~\ref{fig:hist3}).
In other words, by using a dimensionless variable ${X}$ defined as
\begin{eqnarray}	
	X=S/\langle S\rangle_{t}
	\enspace ,
\end{eqnarray}
raw distributions ${\mathrm{P}\left(S;t\right)}$ can be transformed into a time-independent distribution ${\widetilde{\mathrm{P}}\left(X\right)}$ as the following:
\begin{eqnarray}
	\langle S\rangle_{t} \mathrm{P}\left(\frac{S}{\langle S\rangle_{t}};t\right)=\widetilde{\mathrm{P}}\left(X\right).\label{eq:pr2}
\end{eqnarray}
However, this property does not hold in initial stage. 
It is clearly observed in Fig.~\ref{fig:hist2} before ${t=180.0}$.
After the initial stage, Eq.~\eqref{eq:pr2} holds very well with time.

The initial stage of time evolution of $\langle S \rangle_{t}$ corresponds the scaling violation stage. 
By using Eq.~\eqref{eq:cleq}, we estimate the time scale of the initial stage. 
This time scale is determined by the time when ${F\lb t \rb/\Xi \lb t\rb}$ is greater than ${1}$ in Eq.~\eqref{eq:cleq}.
If ${F\lb t \rb/\Xi \lb t\rb}$ is greater than ${1}$, average size decays in proportion to ${\Xi\lb t \rb / F\lb t\rb}$.
As it has been mentioned above, the region of ${\Xi\lb t\rb}$ is limited by yield stress $\sigma_{Y}$.
Furthermore, the lower limit can be estimated as greater than zero because the paste has been shrunk and received tension effectively.
As a rough estimate, supposing that ${\Xi \lb t\rb}$ equals ${\sigma_{Y}}$, the time scale of the initial stage is evaluated to ${\sigma_{Y}/v_{\tau}\sim 227}$.
The actual time scale might be smaller this value, because ${\Xi \lb t\rb}$ is a characteristic stress in the fragments  and smaller than ${\sigma_{Y}}$.

\section{\label{sec:level8}Conclusions and Discussions}
In this article, we modeled a thin drying paste and reproduced crack patterns by means of the smoothed particle hydrodynamics.
Furthermore, we found two properties: The power-law decaying of average size and the scaling law of size distributions by their averages.
According to a dimensional analysis, we found the relationship between the average size and the drying stress.

In our model, since the average size is proportional to inverse time, the number of fragments increases in proportion to time.
This result shows that the drying fracture process is not a simple dividing system such as a cell division whose number of cells increases exponentially.
For understanding this result, it is necessary to consider the dividing process that the dynamics of drying fracture process is considered.
Most of typical distributions, such as normal and log-normal distributions, have two characteristic parameters, an average and a variance, and they can be scaled by using two parameters.
However, the distribution in our model can be scaled by only one parameter, the average.
Therefore, we can consider that these distributions are unknown distributions which can not be described by some typical distributions.
We have not been able to identify the functional form of these distributions yet.
Curz \textit{et al.}\cite{Hidalgo2008} have shown that a mass distribution is scaled by their average mass in a simulation of fragmentation of hard-core granular gases with different restitution coefficients. 
It does not have direct relation with our results. But there are some analogous properties with the present scaled distribution. 

The patterns of our model are similar to the experimental patterns of drying paste.
Therefore, we are required to develop how to measure the similarity of patterns quantitatively and 
confirm the similarity with the experiments. 
We have confirmed only the power decaying and the scaling law of the distribution in experimental results
preliminarily.\cite{priv:2011} In order to give a precise conclusion of both properties, however, 
we must obtain and analyze a lot of experimental data.  
Size distributions of fragments can be scaled by their averages. So if we could know the behavior of average size of fragments, we can know the future distributions of size from a initial distribution. If actual experiments have this scaling law, it is possible to predict the distributions. 

\section*{\label{sec:level9}Acknowledgements}
The authors thank A. Nakahara, S. Kitsunezaki, T. Ooshida and M. Otsuki for useful discussions.
The computation in this work has been done using the facilities of the Supercomputer Center, ISSP, University of Tokyo.
The numerical calculations in this work were carried out on SR16000 at  YITP in Kyoto University.
This work is partly supported by Grant-in-Aid for Scientific Research
(C) No. 22540387 from JSPS, Japan.
S.~I. acknowledges support of the Global COE Program(Core Research and Engineering of Advanced Materials-Interdisciplinary Education Center for Materials Sience).

\end{document}